\documentclass{article}
\usepackage{arxiv}
\usepackage[utf8]{inputenc} 
\usepackage[T1]{fontenc}    
\usepackage{url}            
\usepackage{booktabs}       
\usepackage{amsfonts}       
\usepackage{nicefrac}       
\usepackage{microtype}      
\usepackage{lipsum}		
\usepackage{graphicx}
\usepackage{amsmath}
\usepackage{algorithm}
\usepackage{algorithmicx}
\usepackage[noend]{algpseudocode}
\usepackage{caption,subcaption}
\allowdisplaybreaks
\algnewcommand{\LeftComment}[1]{\Statex \(\triangleright\) #1}
\usepackage{stmaryrd}
\usepackage{color,soul}
\usepackage{wrapfig}
\usepackage{cite}

 \usepackage{palatino}

\usepackage{listings}
\usepackage{xcolor}

\definecolor{codegreen}{rgb}{0,0.6,0}
\definecolor{codegray}{rgb}{0.5,0.5,0.5}
\definecolor{codepurple}{rgb}{0.58,0,0.82}
\definecolor{backcolour}{rgb}{0.95,0.95,0.92}

\lstdefinestyle{mystyle}{
  backgroundcolor=\color{backcolour},   commentstyle=\color{codegreen},
  keywordstyle=\color{magenta},
  numberstyle=\tiny\color{codegray},
  stringstyle=\color{codepurple},
  basicstyle=\ttfamily\footnotesize,
  breakatwhitespace=false,         
  breaklines=true,                 
  captionpos=b,                    
  keepspaces=true,                 
  numbers=left,                    
  numbersep=5pt,                  
  showspaces=false,                
  showstringspaces=false,
  showtabs=false,                  
  tabsize=2
}

\title{Uncertainty Quantification in Friction Model for Earthquakes using Bayesian inference}

\author{
  {
  Saumik Dana}\\
	University of Southern California\\
	Los Angeles, CA 90007 \\
	\texttt{sdana@usc.edu} \\
	\And
  {
  Karthik Reddy Lyathakula} \\
	North Carolina State University\\
	Raleigh, NC 27607 \\
	\texttt{klyatha@ncsu.edu} 
}

\begin{document}
\maketitle
\begin{abstract}
This work presents a framework to inversely quantify uncertainty in the model parameters of the friction model using earthquake data via the Bayesian inference. The forward model is the popular rate- and state- friction (RSF) model along with the spring slider damper idealization. The inverse model is to determine the model parameters using the earthquake data as the response of the RSF model. The conventional solution to the inverse problem is the deterministic parameter values, which may not represent the true value, and quantifying uncertainty in the model parameters increases confidence in the estimation. The uncertainty in the model parameters is estimated by the posterior distribution obtained through the Bayesian inversion.

\end{abstract}
\section{Introduction}

\begin{wrapfigure}{r}{.5\textwidth}
\centering
\includegraphics[trim={0 0 0 0},clip,scale=0.5]{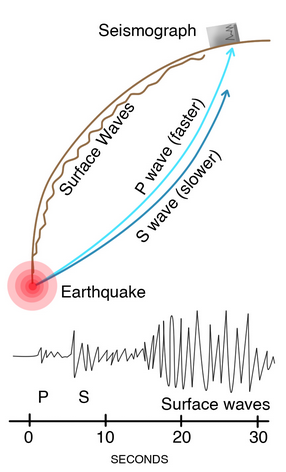}
\caption{P wave arrives first, followed by the S wave and then by surface waves.}
\label{waves}
\end{wrapfigure}

Earthquakes occur as a result of global plate motion. Some earthquakes stop after only a few hundred meters while others continue rupturing for a thousand kilometers. An earthquake produces P waves, or compressional waves, that travel faster and reach the seismograph first, and S waves, or shear waves, that are slower (Fig.~\ref{waves}). Both are transmitted within the Earth and are called body waves. Even slower are surface waves that run along the surface of the earth and do a lot of the damage. The earthquake focus is the point within the Earth where the earthquake originates. The epicenter is a point on the surface directly above the focus. The simplest model for earthquake initiation is to assume that when the stress accumulated in the plates exceeds some failure criterion on a fault plane, an earthquake happens \cite{kanamori2004physics}. Evaluating this criterion requires both a measure of the resolved stress on the fault plane and a
quantifiable model for the failure threshold. The groundbreaking work of \cite{anderson1905dynamics} arrived at the hypothesis that faulting occurs when the resolved shear stress exceeds the internal friction on some plane in the medium leading to fault slip. 

The quantification of earthquakes from the fault slip is achieved using the Rate- and State-dependent Friction (RSF) model, which is considered the gold standard for modeling earthquake cycles (interseismic loading followed by coseismic relaxation) on mature faults \cite{DieJ1979,DieJ1981,RuiA1983,SchC1989,MarC1998}. It is given by
\begin{equation}\label{e:ratestate}
\begin{split}
\mu &= \mu_0 + A\ln{\left(\frac{V}{V_0}\right)} + B\ln{\left(\frac{V_0\theta}{d_c}\right)}, \\
\frac{d\theta}{dt}&= 1-\frac{\theta V}{d_c},
\end{split}
\end{equation}
where $V=|d\boldsymbol{d}/dt|$ is the slip rate magnitude, $a=\frac{dV}{dt}$ which we hypothesize is of the same order as recorded by seismograph, $\mu_0$ is the steady-state friction coefficient at the reference slip rate $V_0$, $A$ and $B$ are empirical dimensionless constants, $\theta$ is the macroscopic variable characterizing state of the surface and $d_c$ is a critical slip distance. Here, $\theta$ may be understood as the frictional contact time \cite{DieJ1979}, or the average maturity of contact asperities between the sliding surfaces \cite{RicJ1993}. The evolution of $\theta$ is assumed to be independent of changes in the normal traction that can accompany the fault slip due to changes in fluid pressure. The model accounts for the decrease in friction (slip-weakening) as the slip increases, and the increase in friction (healing) as the time of contact or slip velocity increases. The two effects act together such that $A>B$ leads to the strengthening of the fault, stable sliding and creeping motion, and $A<B$ leads to weakening of the fault, frictional instability, and accelerating slip. In this way, the model is capable of capturing repetitive stick-slip behavior of faults and the resulting seismic cycle \cite{DieJ1981, SchC1989}. 

The critical slip distance, $d_c$, is the distance over which a fault loses or regains its frictional strength after a perturbation in the loading conditions \cite{palmer1973growth}. In principle, it determines the maximum slip acceleration and radiated energy during an earthquake insofar that it influences the magnitude and time scale of the associated stress breakdown process (e.g., fracture energy) \cite{scholz2019mechanics}. Regardless of the importance, it is paradoxical that the values of $d_c$ reported in the literature range from a few to tens of microns as determined in typical laboratory experiments with bare surfaces and gouge layers \cite{scholz2019mechanics}, to 0.1–5 m as determined in numerical and seismological estimates based on geophysical observations \cite{kaneko2017slip}, and further to several meters as determined in high‐velocity laboratory experiments \cite{niemeijer2011frictional}. Note that among these studies, the critical length scale parameter in the constitutive friction laws ($d_c$) may differ from the slip‐weakening distance inferred from the traction evolution curves obtained for natural or laboratory faults. The latter, as usually derived from scenarios where perturbations are large (e.g., velocity steps of large magnitudes or tips of dynamic rupture nucleation zones), is also referred to as the equivalent or effective slip‐weakening distance ($d_0^{eq}$ or $d_0$, see a review by \cite{marone2009critical}). Moreover, in most numerical simulations of dynamic rupture propagation with prescribed friction laws, $d_c$ is imposed a priori and its value is often assumed to be constant and uniform on the fault plane. Understanding the physics that controls the critical slip distance and explains the gap between observations from experimental and natural faults is thus one of the crucial problems in both the seismology and laboratory communities \cite{ohnaka2003constitutive}. 

With that in mind, we provide a framework in which the earthquake data is used alongside RSF to quantify uncertainty in critical slip distance. While the resolution and coupled flow and geomechanics \cite{dana-2018,dana2019design,dana2019simple,dana2019system,dana2020,dana2020efficient,dana2021,danacg,danacmame,danathesis,dana2021performance} associated with subsurface activity in the realm of energy technologies and concomitant earthquake quantification is a hot topic, in this work, we focus on the effect of stress perturbations in the absence of pore pressure variable. In section 2, we explain the spring slider damper idealization to infer the influence of critical slip distance on RSF without recourse to complicated elastodynamic equations. In section 3, we explain the Bayesian inference framework to inversely quantify uncertainty in the estimation of critical slip distance. In section 4, we present conclusions and outlook for future work.

\section{Spring slider damper idealization to study earthquake response}

We first rewrite Eq. \eqref{e:ratestate} as
\begin{equation}\label{e:ratestate1}
\left.\begin{array}{c}
V = V_0\exp\left(\frac{1}{A}\left(\mu - \mu_0 -B\ln{\left(\frac{V_0\theta}{d_c}\right)}\right)\right),\\
\dot{\theta} = 1-\frac{\theta V}{d_c},\\
\ddot{\theta} = -\frac{\dot{\theta} V}{d_c}
\end{array}\right\}
\end{equation}

\begin{figure}[h]
\centering
\includegraphics[trim={0 0 0 0},clip,scale=0.4]{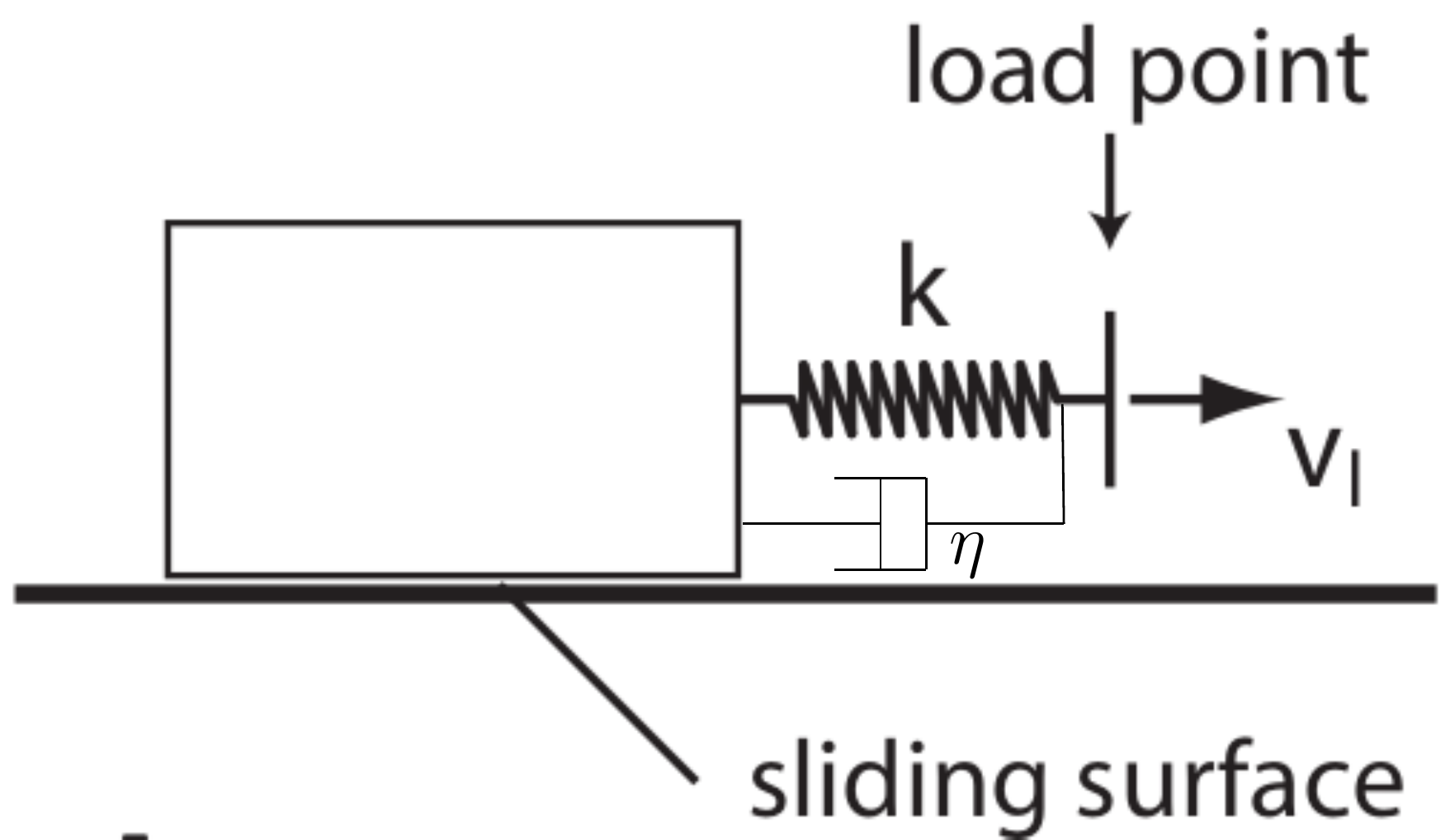}
\caption{Spring Slider Damper Idealization of Fault Behavior}
\label{ss}
\end{figure}

As shown in Fig.~\ref{ss}, we model a fault by a slider spring system \cite{rice1983earthquake,gu1984slip,dieterich1992earthquake}. The slider represents either a fault or a part of the fault that is sliding. The stiffness $k$ represents elastic interactions between the fault patch and the ductile deeper part of the fault, which is assumed to creep at a constant rate. This simple model assumes that slip, stress, and friction law parameters are uniform on the fault patch. 

The friction coefficient of the block is given by
\begin{equation*}\label{slider_spring1}
\mu= \frac{\tau}{\sigma} = \frac{\tau_l-k\delta-\eta V}{\sigma}
\end{equation*}
where $\sigma$ is the normal stress, $\tau$ the shear stress on the interface, $\tau_l$ is the remotely applied stress acting on the fault in the absence of slip, -$k\delta$ is the decrease in stress due to fault slip \cite{kanamori2004physics} and $\eta$ is the radiation damping coefficient \cite{mcclure2011investigation}. We consider the case of a constant stressing rate $\dot{\tau_l} = kV_l$ where $V_l$ is the load point velocity. The initial stress may be smaller or larger than steady state friction owing to coseismic slip on the fault patch or on adjacent parts of the fault. The expression neglects inertia, and is thus only valid for low slip speed in the interseismic period. The stiffness is a function of the fault length $l$ and elastic modulus $E$ as $k\approx \frac{E}{l}$. With $k'=\frac{E}{l\sigma}$, we get
\begin{equation}\label{e:ratestate4}
\left.\begin{array}{c}
\dot{\mu} \approx k'(V_l-V)-k''\dot{V},\\
\ddot{\mu} \approx k'(\dot{V}_l-\dot{V})-k''\ddot{V}
\end{array}\right\}
\end{equation}
where $k''=\frac{\eta}{\sigma}$. Once the phenomenological form of $\dot{\mu}$ and $\ddot{\mu}$ is known, we use the following to get $\dot{V}$ and $\dot{a}$,
\begin{equation}\label{e:ratestate2}
\left.\begin{array}{c}
\dot{V}  = \frac{V}{A}\left(\dot{\mu}-\frac{B}{\theta}\dot{\theta}\right),\\
\dot{a}=\frac{\dot{V}}{A}\left(\dot{\mu}-\frac{B}{\theta}\dot{\theta}\right)+\frac{V}{A}\left(\ddot{\mu}-\frac{B}{\theta}\ddot{\theta}+\frac{B}{\theta^2}\dot{\theta}\right)
\end{array}\right\}
\end{equation}

\subsection{Forward Model Response to a Standard Impulse}

\begin{algorithm}
\caption{Rate and state friction model with radiation damping term}
\begin{algorithmic}
\State Initialize $\theta=\theta_0$, $\mu=\mu_{ref}$
\State Use Eq. \eqref{e:ratestate1} to get $V$, $\dot{\theta}$ and $\ddot{\theta}$
\State Use Eq. \eqref{e:ratestate4} to get $\dot{\mu}$ and $\ddot{\mu}$
\State Use Eq. \eqref{e:ratestate2} to get $\dot{V}$ and $\dot{a}$
\end{algorithmic}
\label{blah}
\end{algorithm}

\begin{figure}[h]
\begin{subfigure}{0.5\textwidth}
	    \centering
	    \includegraphics[scale=0.55]{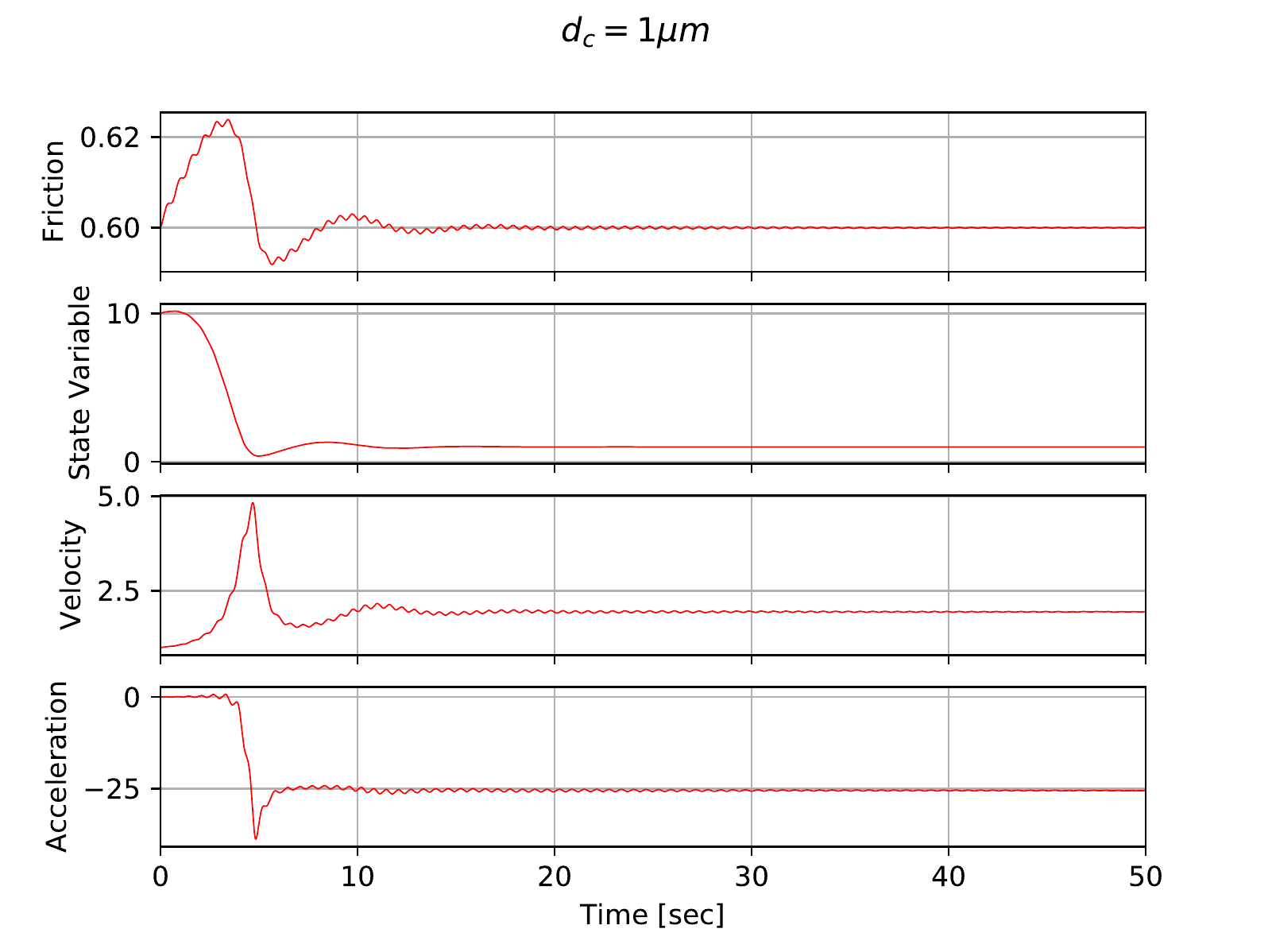}
	    \caption{}
    \end{subfigure}
	\begin{subfigure}{0.5\textwidth}		
		\centering
		\includegraphics[scale=0.55]{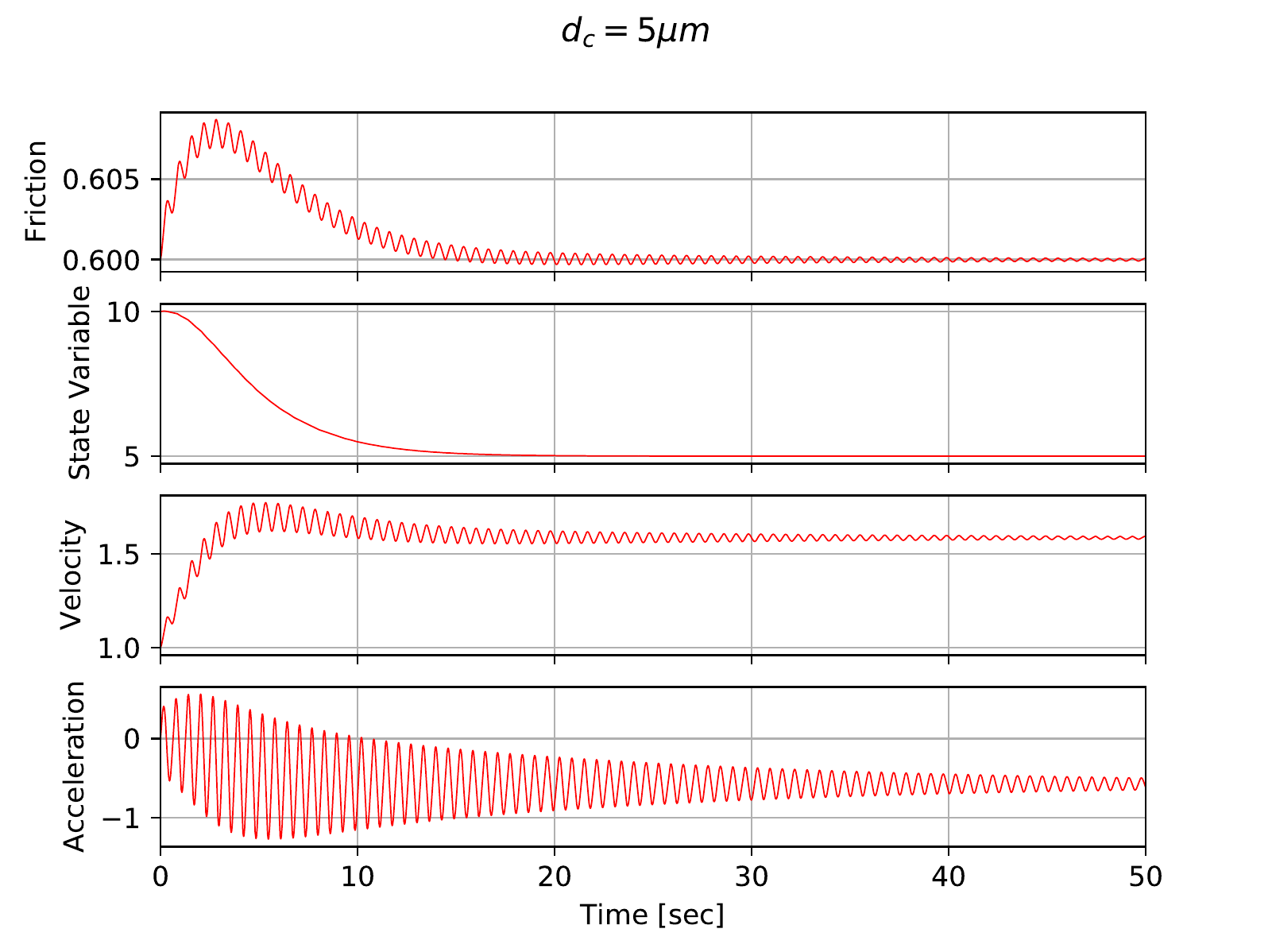}
		\caption{}
	\end{subfigure}
	\begin{subfigure}{0.5\textwidth}
		\centering
		\includegraphics[scale=0.55]{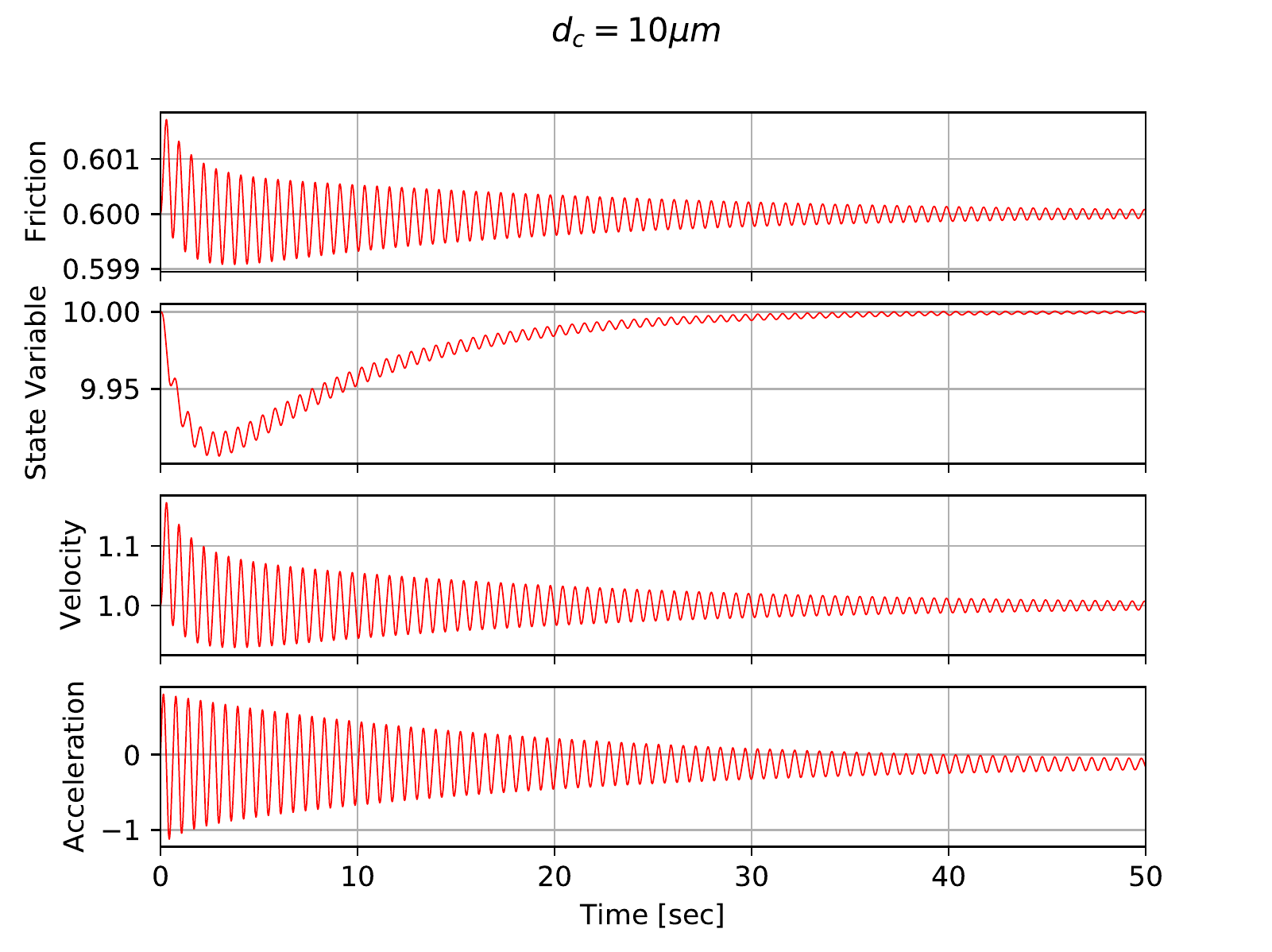}
		\caption{}
	\end{subfigure}
		\begin{subfigure}{0.5\textwidth}
		\centering
		\includegraphics[scale=0.55]{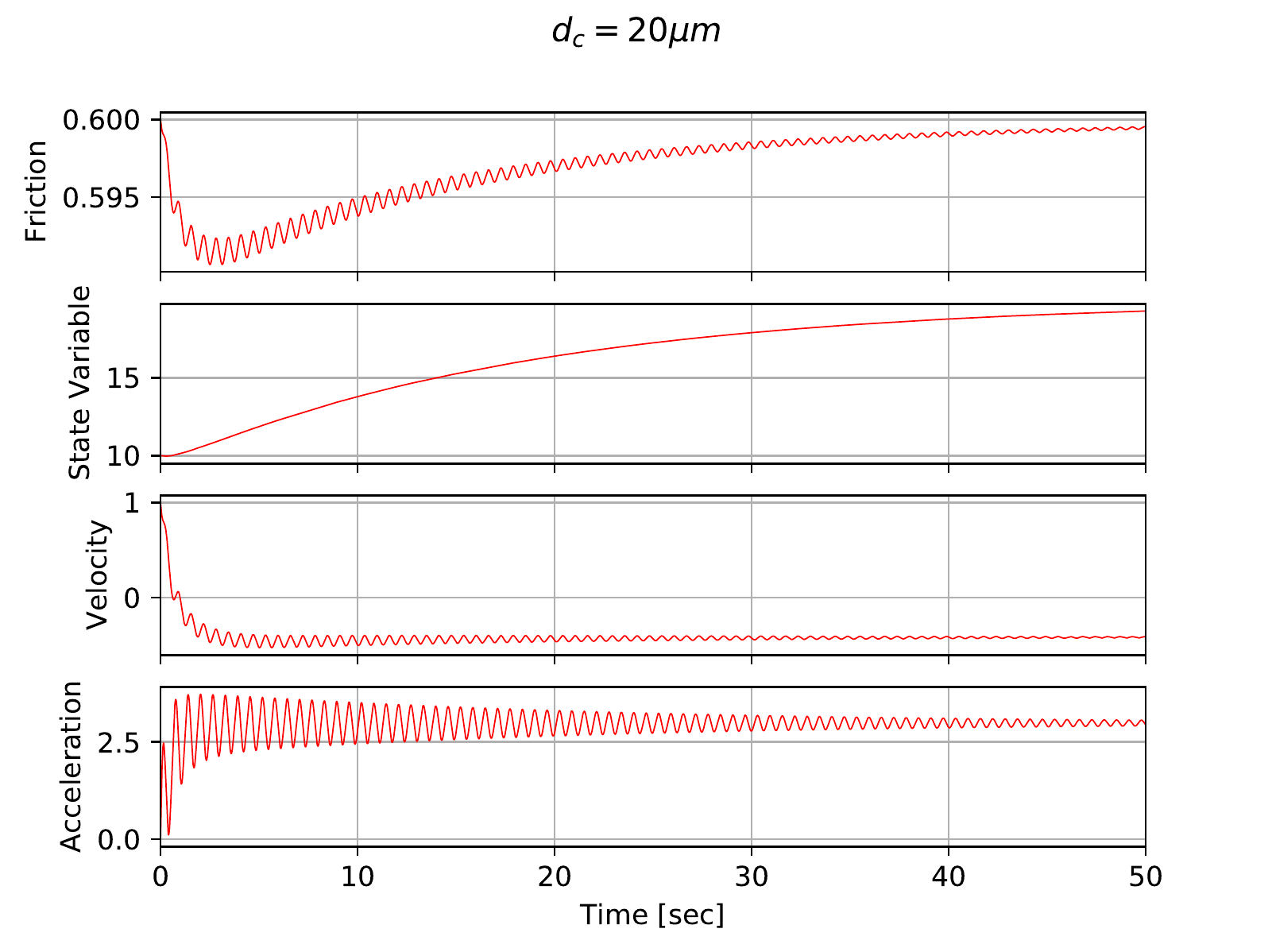}
		\caption{}
	\end{subfigure}
	\caption{System response for different values of critical slip distance. Units of displacement, velocity and accelearation are $\mu m$, $\mu m/s$ and $\mu m/s^2$}
	\label{response}
\end{figure}

We follow the steps outlined in Algorithm \ref{blah} to arrive at the temporal variations of acceleration and fault friction coefficient. We initialize the friction coefficient and state variable and obtain the slip rate and rate of change of the state variable. We then use these values to obtain time derivatives of acceleration and slip rate. These time derivatives are required as we employ the integrated feature of the scientific Python package SciPy \cite{virtanen2020scipy}. The influence of critical slip distance on system response to a load point perturbation of the form
\begin{align*}
V_l = 1+\exp{(-t/20)}\,sin(t/10)
\end{align*}
is shown in Fig.~\ref{response}. The code to generate the plots has been given in Appendix \ref{python}. This code is a part of the GitHub repository \url{https://github.com/karthikncsu/Bayesian-inference-using-earthquake-data}.

The ballpark values are taken from \cite{kanamori2004physics} and \cite{mcclure2011investigation}. Elastic modulus $E=5 \times 10^{10}\,Pa$, Critical fault length $l=3\times 10^{-2}\,m$, Normal stress $\sigma=200 \times 10^6 Pa$, Radiation damping coefficient $\eta=20 \times 10^6 Pa/(m/s)$, $A=0.011$ and $B=0.014$. The effective stiffness and damping is obtained as
\begin{align*}
&k'=\frac{E}{l\sigma}=\frac{5 \times 10^{10}}{3\times 10^{-2}\times 2 \times 10^8} [1/m] \approx 10^{4} [1/m] \equiv 10^{-2} [1/\mu m],\\
&k'' = \frac{\eta}{\sigma} = \frac{2 \times 10^7}{2 \times 10^8} = 0.1 [s/m] \approx 1 \times 10^{-7}[s/\mu m]
\end{align*}

\section{Bayesian Inversion Framework}

\begin{figure}[h]
\centering
\includegraphics[scale=0.6]{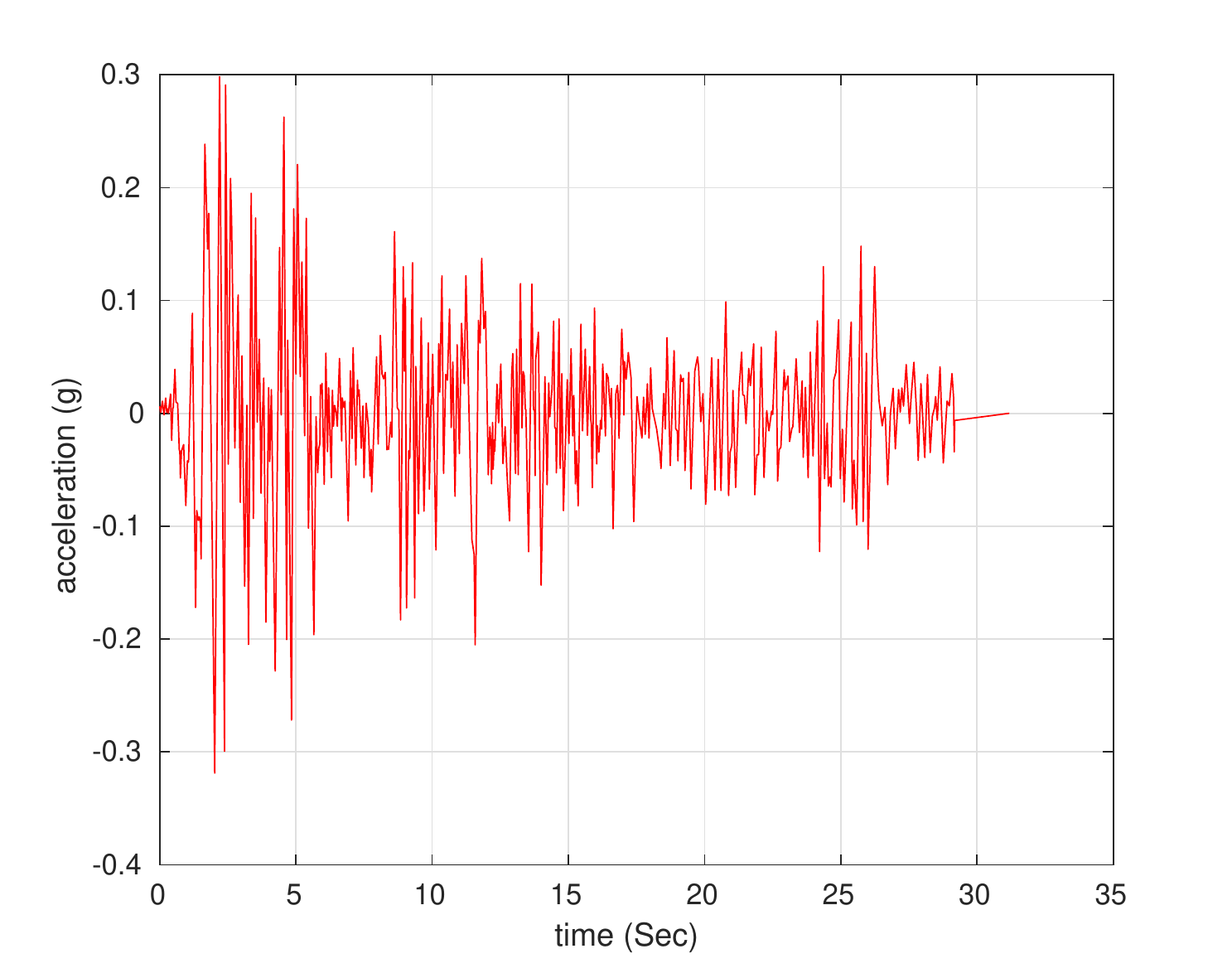}
\caption{Seismograph reading (measured as normalized to $g=9.8\,m/s^2$) of 1940 EL Centro earthquake of magnitude $M_w=6.9$ that occurred in the Imperial Valley in southeastern Southern California near the USA-Mexico border. It was the first major earthquake to be recorded by a strong-motion seismograph located next to a fault rupture, and led to a total damage of $\$6$ million \cite{stover1993seismicity}.}
\label{elcentro}
\end{figure}

Fig.~\ref{response} shows the response of the forward (rate and state friction) model for a given point load perturbation ($V_l$), critical slip distance parameter ($d_c$), and empirical constants ($\mu_0$, $V_0$, $A$ and $B$). The response of the model is the acceleration of the slider, computed using algorithm 1. In an inverse problem, the acceleration response of the model is known and the goal is to find the parameter, critical slip distance parameter ($d_c$). The earthquake acceleration data shown in Fig. \ref{elcentro} can be considered as the acceleration response of the model. To define the inverse problem, considered the relationship between acceleration ($a_i(t)$) and the model response by the following statistical model

\begin{equation}\label{e:ratestate_stat}
\begin{split}
a(t_i) = f(t_i,\theta,\mu,A,B,d_c)+\epsilon_{i}
\end{split}
\end{equation}

where $\epsilon_{i}$ is the error in the statistical model. Here the $a(t_i)$ and $\epsilon_{i}$ are the random variables. The earthquake data over time $a(t_1),...,a(t_n)$ are the $n$ observations for $a(t_i)$ and $f(A,t_i,\theta,\mu,A,B,d_c)$ is the acceleration response of the model over time obtained using the Algorithm 1. The goal of the inverse problem is to determine the model parameter ($d_c$) from the Eq.\eqref{e:ratestate_stat} and conventional method to determine the model parameter that mimizes the norm of the errors using the least squares fit solution as shown below

\begin{equation}\label{e:ratestate_lsfit}
\begin{split}
d_{c,0}= \operatorname*{arg\,min}_{d_c} \sum_{i=1}^{n} (\epsilon_{i})^2 =  \operatorname*{arg\,min}_{d_c} \sum_{i=1}^{n} (a(t_i) - f(t_i,\theta,\mu,A,B,d_c))^2
\end{split}
\end{equation}

The critical slip distance parameter ($d_c$) obtained using the least-squares fit solution, Eq.\eqref{e:ratestate_lsfit} is deterministic value. The values estimated using the least square fit are not the true values due to inherent noise in the data and in most cases, the noise in the data makes it difficult to find the true value. Instead, finding a probability distribution for the model parameters encompasses the true model parameter values and increases the confidence in the prediction. Using the Bayes theorem \cite{smith2013uncertainty}, the distribution for the model parameters is given by the posterior distribution

\begin{equation}\label{e:ratestate_bayes}
\pi(d_c|a(t_1),...,a(t_n))= \frac{\pi(a(t_1),...,a(t_n)|d_c) \pi_0(d_c)}{\int_{d_c} \pi(a(t_1),...,a(t_n)|d_c) \pi_0(d_c) dd_c}
\end{equation}

Here $\pi(d_c|a(t_1),...,a(t_n))$ is the posterior, $\pi(a(t_1),...,a(t_n)|d_c)$ is the likelihood and $\pi_0(d_c)$ is the prior distribution for the model parameters. Assuming the $\epsilon_{i} \sim N(0,\sigma^2)$ as unbiased, independent and identical normal distribution with standard deviation $\sigma$, the likelihood function is expressed as 

\begin{equation}\label{e:ratestate_likeli}
\begin{split}
\pi(a(t_1),...,a(t_n)|d_c) =  \prod_{i=1}^{n} \pi(a(t_i)|d_c) = \prod_{i=1}^{n} \frac{1}{\sigma \sqrt{2\pi}} e^{-\frac{1}{2} \left(\frac{a(t_i)-f(V,t_i,\theta,\mu,A,B,d_c)}{\sigma}\right)^2}
\end{split}
\end{equation}

In above equation, the $f(V,t_i,\theta,\mu,A,B)$ is calculated using the forward problem given by algorithm 1. The information of the model parameters can be included in the posterior distribution through the prior, $\pi_0(d_c)$. In this study, the prior is assumed to be uniform distribution and the prior is a constant value inside the uniform distribution limits.

\begin{algorithm}
\caption{Bayesian inference}
\begin{algorithmic}
\State Input data: earthquake data, $a(t_1),...,a(t_n)$
\State Generate grid $d_c$
\State Use Algorithm 1 to get RHS of Eq. \eqref{e:ratestate_stat} 
\State Use \eqref{e:ratestate_likeli} to get $\pi(a(t_1),...,a(t_n)|d_c)$ for each grid point.
\State Integrate Eq. \eqref{e:ratestate_bayes} to obtain the posterior distribution
\end{algorithmic}
\end{algorithm}

The goal of the inverse problem is to calculate the posterior distribution Eq.\eqref{e:ratestate_bayes}, which represents the uncertainty in the critical slip distance parameter ($d_c$) due to the noise in the earthquake data. Direct evaluation of the posterior distribution using quadrature rules is expensive and often requires adaptive methods to find the posterior distribution. Alternatively, sampling methods like Markov chain Monte Carlo (MCMC) methods \cite{smith2013uncertainty,lyathakula2021fatigue,lyathakula2021probabilistic} can be used to generate samples from the posterior distribution.

\section{Conclusions and Future Work}

This work presents a framework to inversely quantify uncertainty in the critical slip distance parameter of the rate and state friction (RSF) model via the Bayesian inference using the earthquake data. The forward model is to determine the acceleration, using the RSF model, for the given model parameters. In case of an inverse problem, the acceleration data is known and the goal is to find the model parameters. Using conventional methods such as least-squares methods, a deterministic value of the critical slip distance parameter can be obtained from the inverse problem. However, the deterministic parameter value estimated using the conventional methods does not represent the true values due to the noise in the earthquake data, and quantifying uncertainty in the model parameters increases the confidence in the prediction. The uncertainty in the model parameter is estimated by the posterior distribution obtained from the Bayes theorem. The future work will be to demonstrate a simulation to quantify uncertainty in the critical slip distance parameter using the earthquake data via sampling methods such as Markov chain Monte Carlo.  

\clearpage
\appendix
\section{Python Program to Generate Forward Model Response}\label{python}
\begin{lstlisting}[language=Python, caption=Python example]
from scipy.misc import derivative
import numpy as np
from scipy import integrate
import matplotlib.pyplot as plt
from math import exp,log,pi,sin,cos

fig = plt.figure()
fig.suptitle('$d_c=20 \mu m$')

def friction(t,y):
    k = 1e-2
    a = 0.011
    b = 0.014
    Dc = 20.
    mu_ref = 0.6
    V_ref = 1.
    k1 = 1e-7 # radiation damping term
#    k1 = 0

#    if t < 5:
#        V_lp = 1.
#    else:
#        V_lp = 10.

    amp_t = 20
    trig_t = .1
    tp = t/amp_t
    tt = t/trig_t

    amp = 1
    V_lp = 1 + amp*exp(-tp)*sin(tt)
    dV_lp = -amp/amp_t*exp(-tp)*sin(tt) + amp/trig_t*exp(-tp)*cos(tt) # time derivative of V_l
    
    # Just to help readability
    #y[0] is mu (friction)
    #y[1] is theta
    #y[2] is velocity
    #y[3] is acceleration

    n = len(y)
    dydt = np.zeros((n,1))

    # compute v
    temp_ = V_ref * y[1] / Dc
    temp = 1/a*(y[0] - mu_ref - b * log(temp_))
    v = V_ref * exp(temp)

    # time derivative of theta
    dydt[1] = 1. - v * y[1] / Dc

    # double derivative of theta
    ddtheta = - dydt[1]*v/ Dc

    # time derivative of mu
    dydt[0] = k*V_lp - k*v

    # time derivative of velocity
    dydt[2] = v/a*(dydt[0] - b/y[1]*dydt[1])

    # double derivative of mu
    ddmu = k*dV_lp - k*dydt[2]

    # time derivative of acceleration
    dydt[3] = dydt[2]/a*(dydt[0] - b/y[1]*dydt[1]) + v/a*(ddmu - b/y[1]*ddtheta + b/y[1]*dydt[1]/y[1])

    # radiation damping
    dydt[0] = dydt[0] - k1*dydt[2]
    dydt[2] = v/a*(dydt[0] - b/y[1]*dydt[1])
    ddmu = ddmu - k1*dydt[3]
    dydt[3] = dydt[2]/a*(dydt[0] - b/y[1]*dydt[1]) + v/a*(ddmu - b/y[1]*ddtheta + b/y[1]*dydt[1]/y[1])
    
    return dydt

r = integrate.ode(friction).set_integrator('vode', order=5,max_step=0.001,method='bdf',atol=1e-10,rtol=1e-6)

# Time range
t_start = 0.0
t_final = 50.
delta_t = 1e-2
num_steps = int(np.floor((t_final-t_start)/delta_t)+1)

# Initial conditions
mu_t_zero = 0.6
V_ref = 1.
Dc = 10.
mu_ref = 0.6
theta_t_zero = Dc/V_ref
v = V_ref
start_acc = 0
r.set_initial_value([mu_t_zero, theta_t_zero, V_ref, start_acc], t_start)

# Create arrays to store trajectory
t = np.zeros((num_steps,1))
mu = np.zeros((num_steps,1))
theta = np.zeros((num_steps,1))
velocity = np.zeros((num_steps,1))
acc = np.zeros((num_steps,1))
t[0] = t_start
mu[0] = mu_ref
theta[0] = theta_t_zero
velocity[0] = v
acc[0] = 0

# Integrate the ODE(s) across each delta_t timestep
k = 1
while r.successful() and k < num_steps:
    #integrate.ode.set_f_params(r,velocity,k)
    r.integrate(r.t + delta_t)

    # Store the results to plot later
    t[k] = r.t
    mu[k] = r.y[0]
    theta[k] = r.y[1]
    velocity[k] = r.y[2]
    acc[k] = r.y[3]
    k += 1

# Make some plots

#cjm_t,cjm_mu = np.loadtxt('cjm_step.tim',skiprows=2,unpack=True)

ax1 = plt.subplot(411)
ax1.plot(t, mu,color='r', linewidth=0.5)
ax1.set_xlim(t_start, t_final)
#ax1.plot(cjm_t+10.,cjm_mu,color='k')
#ax1.set_xlabel('Time [sec]')
ax1.set_xticklabels([])
ax1.set_ylabel('Friction')
ax1.grid('on')

ax2 = plt.subplot(412)
ax2.plot(t, theta, 'r', linewidth=0.5)
ax2.set_xlim(t_start, t_final)
#ax2.set_xlabel('Time [sec]')
ax2.set_ylabel('State Variable')
ax2.set_xticklabels([])
ax2.grid('on')

ax3 = plt.subplot(413)
ax3.plot(t, velocity, 'r', linewidth=0.5)
ax3.set_xlim(t_start, t_final)
#ax3.set_xlabel('Time [sec]')
ax3.set_ylabel('Velocity')
ax3.set_xticklabels([])
ax3.grid('on')

ax4 = plt.subplot(414)
ax4.plot(t, acc, 'r', linewidth=0.5)
ax4.set_xlim(t_start, t_final)
ax4.set_xlabel('Time [sec]')
ax4.set_ylabel('Acceleration')
ax4.grid('on')

#ax5 = plt.subplot(515)
#ax5.plot(t, forcing, 'r')
#ax5.set_xlim(t_start, t_final)
#ax5.set_xlabel('Time [sec]')
#ax5.set_ylabel('Forcing')
#ax5.grid('on')

plt.show()

\end{lstlisting}

\bibliographystyle{unsrt}     
\bibliography{sample}   
\end{document}